\documentclass[superscriptaddress,prb,twocolumn,tightenlines,balancelastpage,10pt,a4paper]{revtex4}%
\usepackage{bbm}
\usepackage{amsfonts}
\usepackage{amssymb}
\usepackage{graphicx}
\usepackage{subfigure}
\usepackage{savesym}
\usepackage{amsmath}
\usepackage{txfonts}
\usepackage{multirow}
\usepackage{epstopdf}

\newcommand{\ket}[1]{\vert#1\rangle}

\usepackage{soul}

\usepackage{color}

 \savesymbol{iint}
\restoresymbol{TXF}{iint}

\newcommand{\changeda}[1] {\textcolor{black}{#1}}
\newcommand{\changedb}[1] {\textcolor{black}{#1}}

\newcommand{\changedg}[1] {\textcolor{black}{#1}}

\begin{document}

\author{Xiao-Qi Zhou}
\affiliation{Centre for Quantum Photonics, H. H. Wills Physics
Laboratory \& Department of Electrical and Electronic Engineering,
University of Bristol, BS8 1UB, United Kingdom}
\author{Timothy C. Ralph}
\affiliation{Department of Physics and Centre for Quantum Computation and Communication Technology, University of Queensland, Brisbane 4072, Australia}
\author{Pruet Kalasuwan}
\affiliation{Centre for Quantum Photonics, H. H. Wills Physics
Laboratory \& Department of Electrical and Electronic Engineering,
University of Bristol, BS8 1UB, United Kingdom}
\author{Mian Zhang}
\affiliation{Centre for Quantum Photonics, H. H. Wills Physics
Laboratory \& Department of Electrical and Electronic Engineering,
University of Bristol, BS8 1UB, United Kingdom}
\affiliation{School of Applied and Engineering Physics, Cornell University, Ithaca, NY 14853}
\author{Alberto Peruzzo}
\affiliation{Centre for Quantum Photonics, H. H. Wills Physics
Laboratory \& Department of Electrical and Electronic Engineering,
University of Bristol, BS8 1UB, United Kingdom}
\author{Benjamin P. Lanyon}
\affiliation{Institut f\"{u}r Experimentalphysik, Universit\"{a}t Innsbruck, Technikerstr. 25, 6020 Innsbruck, Austria \vspace{-4pt}}
\author{Jeremy L. O'Brien}
\email{Jeremy.OBrien@bristol.ac.uk}
\affiliation{Centre for Quantum Photonics, H. H. Wills Physics
Laboratory \& Department of Electrical and Electronic Engineering,
University of Bristol, BS8 1UB, United Kingdom}

\title{{{Adding control to arbitrary unknown quantum operations}}\vspace{-6pt}}

\begin{abstract}
\changedg{While quantum computers promise significant advantages, the complexity of quantum algorithms remains a major technological obstacle. We have developed and demonstrated an architecture-independent technique that simplifies adding control qubits to arbitrary quantum operations---a requirement in many quantum algorithms, simulations and metrology. The technique is independent of how the operation is done, does not require knowledge of what the operation is, and largely separates the problems of how to implement a quantum operation in the laboratory and how to add a control. We demonstrate an entanglement-based version in a photonic system, realizing a range of different two-qubit gates with high fidelity.}

\end{abstract}

\maketitle
\noindent


Perhaps the most promising future application of quantum science is quantum information processing, which promises secure communication \cite{gi-rmp-74-145} and greatly increased speeds for solving certain problems such as database searching \cite{grover97}, factoring \cite{sh-conf-94-124} and quantum simulation \cite{fe-ijtp-82-467}. The excitement surrounding quantum computers lies in the fact that the number of elementary operations that they require to solve these problems scales only polynomially with the size of the input, in contrast to exponential scaling on a conventional computer. However, even a polynomial scaling of quantum computational resources still presents an enormous obstacle to practical realization. It may well be that although a quantum computer could in principle efficiently solve these important problems, it will remain practically infeasible to build one that can implement a sufficient number of operations, on enough qubits and with sufficient precision, to do anything useful. This motivates developing methods to reduce the resource overhead required to implement key quantum algorithms.

Quantum algorithms rely on the decomposition of a functional quantum circuit into an elementary logic gate set, such as that formed by single-qubit and two-qubit controlled-NOT (CNOT) gates \cite{nielsen}; there have been several experiments to demonstrate universal quantum gate sets in different physical architectures \cite{la-nat-464-45} including ion traps \cite{sc-nat-422-408,le-nat-422-412}, linear optics \cite{pi-pra-68-032316,ob-nat-426-264,ga-prl-93-020504,ba-prl-98-170502,ga-prl-104-020501}, superconductors \cite{st-sci-313-1423,pl-nat-447-836}, and atoms \cite{ma-nat-425-937,an-nat-448-452}, and even small-scale algorithms \cite{la-prl-99-250505,lu-prl-99-250504,po-sci-325-1221,la-nchem-2-106}. However, the large number of elementary gates required to implement even modest sized circuits presents a significant challenge. This complexity is due not only to the sheer number of elementary operations required but also the structure in which these gates are combined.


\changedg{Controlled-unitary ($CU$) gates are a particularly important class of circuits, where one ``control" qubit turns on or off a unitary operation $U$ acting on a register of ``target" qubits (Fig. 1a).} These circuits features heavily in Kitaev's phase estimation algorithm \cite{kitaev2002classical} which underpins Shor's factoring algorithm \cite{sh-conf-94-124} and quantum simulation \cite{fe-ijtp-82-467}. In the context of quantum simulation, $U$ could represent a simulation of the time-evolution operator of some physical system and the ability to add control qubits allows energy eigenvalues to be read out via the phase estimation algorithm \cite{as-sci-309-1704}. Phase estimation is also a fundamental tool in quantum metrology. However, the current standard method of realizing $CU$ gates, which relies on the decomposition of $U$ into an elementary gate set, may not be suitable for these applications: in Kitaev's phase estimation algorithm $U$ may be an unknown ``black box" which cannot be decomposed at all.


\changedg{Here we present and demonstrate a simple method for realizing controlled quantum operations ($CO$), of which $CU$ gates are a subset.}
\changedg{In the following we first explain the technique in a way that is independent of any particular physical system, before describing a conceptual example in a photonic system. We then show that the equivalent operation can be achieved by exploiting an entangled initial state---reminiscent, but distinct from, cluster state quantum computing \cite{ra-prl-86-5188,wa-nat-434-169}. 
Finally, we apply this approach in a series of proof-of-principle experiments implementing various two-qubit gates which include a CNOT, a number of other $CU$ gates, as well as  ``entanglement filter" and ``entanglement splitter" gates.}

\vspace{5mm}

\noindent\textbf{\large{\changedb{Results}}}\\

\begin{figure}[t!]
\begin{center}
\includegraphics[width=\columnwidth]{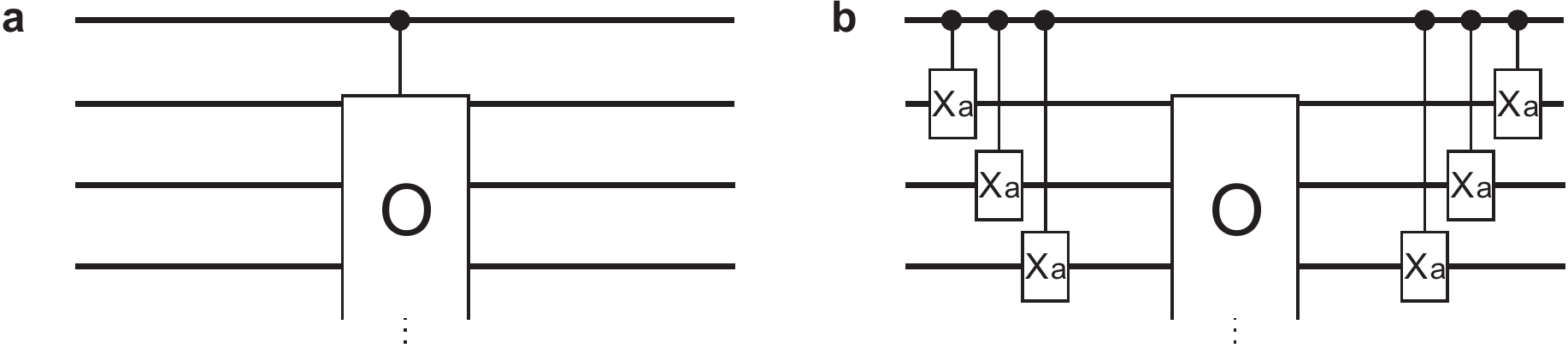}
\end{center}
\par
\vspace{-0.5cm}
\caption{\textbf{Controlling arbitrary quantum operations using additional degrees of freedom.} (\textbf{a})
Logic circuit in which quantum operation $O$ is implemented on a register of qubits (target register), conditional on the logical state of a single control qubit.
(\textbf{b}) Our approach to implementing the circuit in (a). The target information carriers are four dimensional systems with logical states $\ket{0},\ket{1},\ket{2}$ and $\ket{3}$. Initially and finally only the bottom two `qubit' levels ($\ket{0}$ and $\ket{1}$) are populated. Controlled-$X_a$ gates (see Equation~\ref{eq1} and \ref{eq2}) swap information between the qubit levels and the upper levels ($\ket{2}$ and $\ket{3}$), on which $O$ does not act. In this way, conditional on the state of the control qubit, the entire quantum state of the target register is temporarily moved into an effective quantum memory on which $O$ does not act.}
\vspace{-0.5cm}
\label{fig1}
\end{figure}

\noindent\textbf{\changedb{Explanation of the general technique.}}
Our approach for adding a control qubit to an arbitrary quantum operation $O$ is shown conceptually in Figure~\ref{fig1}b. In summary, conditional on the logical state of the control, the quantum state of the target register ($\psi$) is temporarily shifted into a part of an extended Hilbert space on which $O$ does not act (imparts the identity operation). In this way the evolution of $\psi$ is dependent on the control qubit state: if it is $\ket{0}$ ($\ket{1}$) then $\psi{\rightarrow}\psi$ ($\psi{\rightarrow}O\psi$). The Hilbert space is extended by employing an extra two levels in each quantum information carrier in the target register, making each a four level system with logical states $\ket{0},\ket{1},\ket{2}$ and $\ket{3}$. 
The action of each $X_a$ gate is to swap information between the bottom two `qubit' levels ($\vert 0\rangle$, $\vert 1\rangle$)  and the expanded Hilbert space ($\vert 2\rangle$, $\vert 3\rangle$) i.e.
\begin{equation}
X_a=\left(
\begin{array}{cccc}
0 & 0 & 1 & 0 \\
0 & 0 & 0& 1 \\
1 & 0 & 0& 0 \\
0 & 1 & 0 & 0
\end{array}\right)
\label{eq1}
\end{equation}
\begin{equation}
\begin{split}
X_a\vert 0\rangle = \vert 2\rangle,&\quad\quad X_a\vert 1\rangle = \vert 3\rangle\\
X_a\vert 2\rangle = \vert 0\rangle,&\quad\quad X_a\vert 3\rangle = \vert 1\rangle
\end{split}
\label{eq2}
\end{equation}




In spite of its conceptually simplicity, the technique has significant practical benefits: it largely separates the experimental problems of how to implement any given quantum operation in the laboratory and how to add a control qubit. This is relevant in many experimental cases where it is not at all clear how to directly add a control to a quantum operation, for example when $O$ can be realized in analog fashion by turning on an experimental hamiltonian, or when $O$ is a non-unitary operation implemented by directly coupling to a bath. Even in the case where $O$ can be constructed with a universal gate set, the number of additional operations required to add a control will generally be far less following our approach. Furthermore, in situations where $O$ is unknown our method may be the only way to add control. This is relevant in quantum metrology where the goal is to measure properties of $O$. 

The method can be straightforwardly extended to realize the conditional implementation of two different operations $O_1$ or $O_2$ based on the state of the control qubit.  
Here, while the component of the state that is unmoved undergoes $O_1$ , the component of the state moved into the expanded Hilbert space undergoes $O_2$. 
A further extension would be to add multiple control qubits to implement one of several quantum operations, based on the state of all of the control qubits. For example with two control qubits four operations $O_1$, $O_2$, $O_3$, or $O_4$ could be implemented depending on the state of the control qubits.


An alternative approach to extending the Hilbert space would be to use another register of qubits and controlled-swap operations to move information between the registers. However, while adding more qubits has proved to be a significant experimental challenge, multi-dimensional quantum information carriers are readily available in most physical systems currently being investigate or used for quantum information processing. Trapped ions systems, for example, offer a large number of precisely controllable internal electronic and external vibrational degrees of freedom. Our technique could be implemented by conditionally moving quantum information between different electronic transitions, on which subsequent operations do not operate, or operate differently, on. 

\changedg{We note that in previous work it was shown how moving part of the state of a target qubit into an expanded Hilbert space can simplify adding control qubits~\cite{la-nphys-5-134}. However, this only works in the case where the target is a single-qubit unitary and is at the expense of changing how the unitary must be implemented.}

\begin{figure*}[t]
\begin{center}
\includegraphics[width=0.8\textwidth]{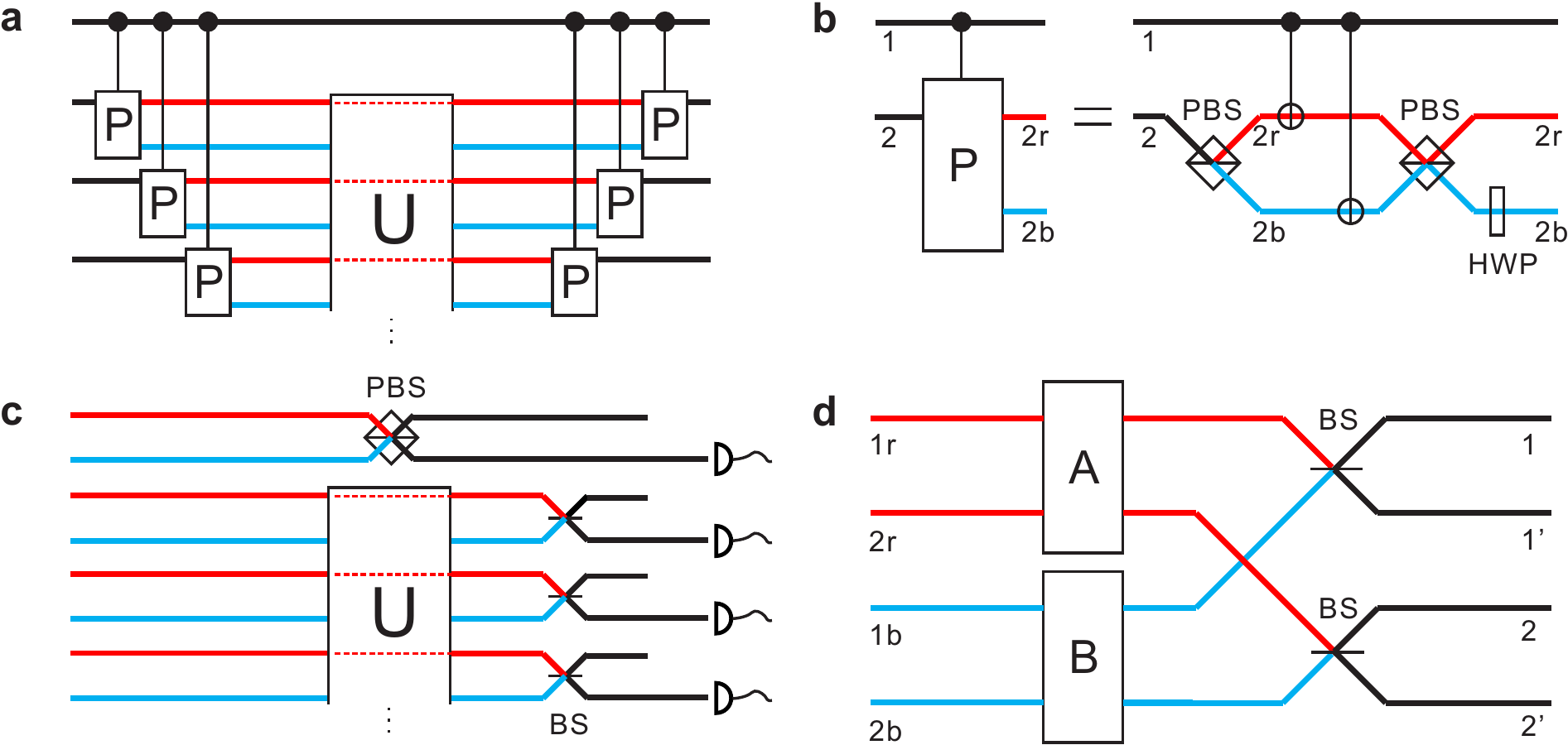}
\end{center}
\vspace{-0.5cm}
\caption{{\textbf{Optical implementation of the scheme.}
(\textbf{a}) \changedb{Implementation of a $CU$ gate. The circuit is composed of
three parts: controlled-path gates ($CP$) at the beginning, which move
the target qubits from upper spatial modes (labeled red) to lower spatial modes (labeled blue) if
the control qubit is in the logical state $\vert 1 \rangle$; the unitary gate $U$ in
the middle is implemented on the lower spatial modes; and  $CP$s at
the end, which combine the upper and lower spatial modes.}
\changedb{(\textbf{b}) Construction of a $CP$ gate. 
When the control is in the logical state $\vert 0 \rangle$, nothing is applied on the target and the photon exits in the red mode. When the control is $\vert 1 \rangle$, the polarization of the target flips and the photon exits in the blue mode. The half waveplate (HWP) flips back the polarization of the target photon in the blue mode.}
(\textbf{c}) Entanglement-based scheme. Instead of implementing $CP$s before the unitary gate $U$, one set the input state in an equal superposition of photons either all in the red modes or all in the blue modes. The $CP$s after $U$ are substituted by a series of non-polarization beamsplitters (BS).
(\textbf{d}) \changedb{Linear combinations of quantum operations}}.
\changedb{The two photons are either in the red modes $1r, 2r$ or in the blue modes $1b, 2b$  which pass through quantum operations $A$ and $B$ respectively. Then two spatial modes $1r, 1b$ ($2r, 2b$) of photon 1 (2) are combined at a BS. By postselecting two photons at ports $1, 2$ or $1', 2'$ ($1, 2'$ or $1', 2$), one can effectively realize the quantum operation $A+B$ ($A-B$).} }
\vspace{-0.5cm}
\label{fig2}
\end{figure*}

\vspace{5mm}
\noindent\textbf{\changedb{Optical version of the scheme}}
Although our technique is independent of the particular physical system and degree of freedom employed, it is particularly well suited to an optical version in terms of the polarization and spatial degrees of freedom of photonic qubits. 
As shown in Figure~\ref{fig2}a, the controlled-path ($CP$) gate substitutes the $CX_a$ in Figure~\ref{fig1}b. \changedg{The $CP$ is a two-photon gate that changes the target photon's path if the control is vertically polarized.} 
We note that the $CP$ gate has previously been proposed for implementing controlled gates in the context of weak optical cross-Kerr non-linearities\cite{ql-pra-80-042310,ql-pra-80-042311}.

\changedb{To understand how the $CP$ gate works, let's examine its structure shown in Figure~\ref{fig2}b. Assume the inputs are two polarization-encoded photonic qubits, $\alpha |H \rangle_1 + \beta |V \rangle_1$ (control photon $1$) and $\gamma |H \rangle_2 + \delta |V \rangle_2$ (target photon $2$). The first PBS will convert the target state from $\gamma |H \rangle_2 + \delta |V \rangle_2$ to $\gamma |H \rangle_{2b} + \delta |V \rangle_{2r}$ where $2r$ and $2b$ denote the red and blue spatial modes of the target photon respectively. The subsequent two CNOT gates flip the polarization of the target photon if the first photon is vertically polarized.  (It is assumed that if a CNOT acts on an unoccupied spatial mode the identity is enacted.) Thus, the two-photon state becomes $\alpha |H \rangle_1(\gamma |H \rangle_{2b} + \delta |V \rangle_{2r})+ \beta |V \rangle_1(\gamma |V \rangle_{2b} + \delta |H \rangle_{2r})$. Then the two spatial modes $2r$ and $2b$ of the target photon are mixed on the second PBS which converts the two-photon state to $\alpha |H \rangle_1(\gamma |H \rangle_{2r} + \delta |V \rangle_{2r}) + \beta |V \rangle_1(\gamma |V \rangle_{2b} + \delta |H \rangle_{2b})$. Finally, a half-waveplate flips the polarization in spatial mode $2b$ and thus converts the state to $\alpha |H \rangle_1(\gamma |H \rangle_{2r} + \delta |V \rangle_{2r}) + \beta |V \rangle_1(\gamma |H \rangle_{2b} + \delta |V \rangle_{2b})$\changedg{: the result is that the target polarisation qubit is to be found in one of two orthogonal spatial modes, depending on the logical state of the control qubit.}} By defining $|H \rangle_{2b}$, $|V \rangle_{2b}$, $|H \rangle_{2r}$ and $|V \rangle_{2r}$ as $\vert 0\rangle$, $\vert 1\rangle$, $\vert 2\rangle$ and $\vert 3\rangle$ respectively, one can easily find that a $CP$ exactly realizes the function of a $CX_a$ gate.

\changedg{Returning to Fig.~\ref{fig2}a, suppose that the control photon is again initially in the arbitrary polarization-qubit state $\alpha |H \rangle + \beta |V \rangle$ and the target photons are initially in the multi-qubit state $|\psi \rangle$. The photons pass through a sequence of CPs which changes the path of all target photons if the control photon is vertically polarized, thus the state is converted to $\alpha |H \rangle|\psi\rangle_r + \beta |V \rangle|\psi\rangle_b$ where $r$ and $b$ denote the collective red and blue spatial
modes respectively. Next the blue spatial modes $b$ of the target photons are acted upon by $U$, while the red spatial modes $r$ do not pass through the unitary, as indicated by the dotted lines. The state is therefore converted to $\alpha |H \rangle |\psi \rangle_r + \beta |V \rangle U |\psi \rangle_b$. Finally, by repeating the sequence of CPs we obtain the desired state $\alpha |H \rangle |\psi \rangle + \beta |V \rangle U |\psi \rangle$ at the output.}


\changeda{There is a clear advantage over the conventional quantum computational approach to adding control qubits, in terms of the number of logic gates required.  Assuming that $U$, which is a unitary acting on n qubits, can be decomposed into a circuit of $p$ CNOTs and $q$ single-qubit gates, one would use $p$ Toffoli gates and $q$ two-qubit controlled gates to build the corresponding CU gate, which can further be decomposed into $(3p+q)$ to $(6p+2q)$ CNOTs and  even more single-qubit gates \cite{la-nphys-5-134}. While $4n$ more CNOTs suffice to add a control to the n-qubit unitary $U$ by using our method, at least $(2p+q)$ more CNOTs are needed if one sticks to the traditional scheme. For most quantum algorithm applications, such as Shor's algorithm where $U$ is a modular exponentiation gate, the typical value of $(p+q)$ is about $72n^3$ which is on the order of $O(n^3)$ and thus $(2p+q)$ is much larger than $4n$ when $n$ becomes large \cite{vl-pra-54-147,be-pra-54-1034}. Furthermore, if $U$ is a non-unitary operation then the conventional approach is to rewrite this as a unitary on a larger Hilbert space, at a significant cost to the number of gates required. There is no additional cost to add controls to non-unitary operations following our approach.}



\begin{figure*}[t]
\begin{center}
\vspace{-0.5cm}
\includegraphics[width=1\textwidth]{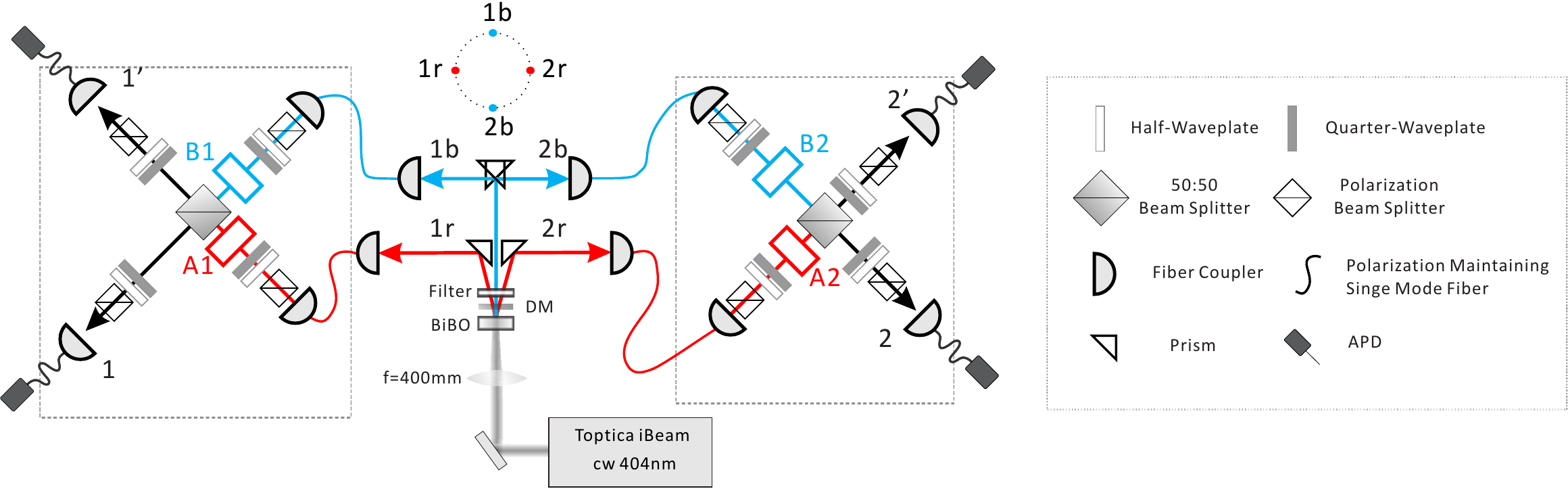}
\vspace{-0.5cm}
\end{center}
\caption{\textbf{Experimental setup for realizing CU gates.} A \changedb{60 mW} continuous-wave (CW) laser beam with a central wavelength of \changedb{404}nm
is focused \changedb{onto} a BiBO crystal to create photon-pairs. Both the horizontal (modes 1r and 2r) and vertical (modes 1b and 2b) photon pairs are collected. Prior to collection into polarization-maintaining fibers (PMF), the photons are spectrally filtered by narrow-band filters ($\Delta\lambda_{FWHW} = 3.2$ nm). $A1$, $B1$, $A2$ and $B2$ are four single-qubit gates. By post-selecting the case the two photons exit at ports $1$ and $2$, one would effectively realize a two-qubit quantum gate $(A1\otimes A2+B1\otimes B2)$. The phase between the two components is stabilized by monitoring the coincidence count rates between detectors \changedb{$1'$ and $2'$}. 
}
\label{fig3}
\vspace{-0.5cm}
\end{figure*}

\vspace{5mm}

\noindent\textbf{Entanglement-based scheme}
\changedg{Even a two-qubit demonstration of our scheme would require two $CP$ gates and subsequently four CNOT gates, which is presently out of reach using current linear-optical quantum information processing technology.} 
\changedg{However, the effect of the $CP$ gate is to generate entanglement between the control qubit and target register, and this kind of entanglement can be generated directly from existing photon sources.
We now present an alternative entanglement-based version of our scheme which is feasible with current technology. 
}

Consider the optical circuit schematic shown in Figure~\ref{fig2}c. At the input a spatially-entangled $n$-photon state is injected that is an equal superposition of finding one photon in each of the $n$ red modes and one photon in each of the $n$ blue modes. In the case where the polarization of the $n$-photon input state is $|\phi\rangle=(\alpha|H\rangle{+}\beta|V\rangle)|\psi\rangle$, where $\alpha |H \rangle + \beta |V \rangle$ is the upper control qubit and $|\psi \rangle$ is the joint state of the lower $(n-1)$ target register qubits, then the initial state can be written as
\begin{eqnarray}
\frac{1}{\sqrt{2}} (|\phi \rangle_r |vac\rangle_b^{ \otimes n}+ |vac\rangle_r^{ \otimes n}|\phi\rangle_b)
 \label{A}
\end{eqnarray}
where $r$ and $b$ label the collective red and blue spatial modes respectively, and $|vac\rangle$ represents an unoccupied mode. From now on we will drop the unoccupied vacuum modes from the notation. Note that, as we will show, it is possible to create such a state from a spontaneous parametric down conversion photon source. The quantum operation $U$ acts only on photons in the blue spatial modes of the target register. The information about whether the target state $\psi$ does or does not undergo the operation $U$ is therefore encoded in the spatial mode of the control photon.
Mixing the two control modes on a polarizing beamsplitter (PBS) and post-selecting on finding the control photon in the lower spatial mode moves this information into the polarization of the control photon, yielding the state:
%
\begin{eqnarray}
\frac{1}{\sqrt{2}}(\alpha |H \rangle |\psi \rangle_r+ \beta |V
\rangle U |\psi \rangle_b)
\label{B}
\end{eqnarray}
Finally, the red and blue modes of each target qubit are mixed on non-polarisaing beamsplitters (BS) to remove the path information. In the case where all photons exit in the lower paths, the output state is
\begin{eqnarray}
\frac{1}{\sqrt{2^{n}}}(\alpha |H \rangle |\psi \rangle+ \beta |V
\rangle U |\psi \rangle) \label{C}
\end{eqnarray}
as required. The probability of success is $(1/2)^{n}$, however all combinations of the control photon arriving in lower spatial mode and an even number of target photons arriving in lower spatial modes will give a state with the same form as in Eq.\ref{C}. There are $2^{n-2}$ such combinations so the total probability of success is $1/4$, regardless of the number of qubits $U$ acts on. The important feature of our approach --- that any operation~(known or unknown) can be controlled without changing the way the operation is done --- is retained in the entanglement based approach.

This approach can be reformulated in a more general way as shown in Figure~\ref{fig2}d. Here we consider the two-photon case for simplicity. Beginning with the two photon input state of Equation~\ref{A}, the red and blue modes pass through quantum operations $A$ and $B$ respectively. The state $\frac{1}{\sqrt{2}}(A|\phi \rangle_r + B|\phi\rangle_b)$ is obtained. After mixing the spatial modes on the two BSs, one would get $(A+B)|\phi\rangle$ if the two photons exit at ports $1$ and $2$ or $1'$ and $2'$. Otherwise, if the two photons exit at ports $1$ and $2'$ or $1'$ and $2$, $(A-B)|\phi\rangle$ would be obtained. 
To realize the $CU$ gate, one just needs to set $A=\vert H\rangle\langle H\vert\otimes I$ and $B=\vert V\rangle\langle V\vert\otimes U$, where $\vert H\rangle\langle H\vert$ and $\vert V\rangle\langle V\vert$ denote projectors onto $\vert H\rangle$ and $\vert V\rangle$ respectively.

This approach provides a new perspective on constructing quantum gates: While the traditional decomposition method can be regarded as performing multiplication, which corresponds to rewriting the target gate matrix as the product of several gate matrices, our method is performing linear combination which means rewriting the target gate matrix as the sum of several gate matrices. 
This entanglement-based scheme would be useful for small scale applications as well as subroutines in larger calculations, and could be made deterministic following the original prescription for linear optical quantum computation\cite{kn-nat-409-46}. Introducing non-linearities is another way to approach a unit success probability~\cite{ql-pra-80-042310}.


\begin{figure*}[t]
\begin{center}
\includegraphics[width=\textwidth]{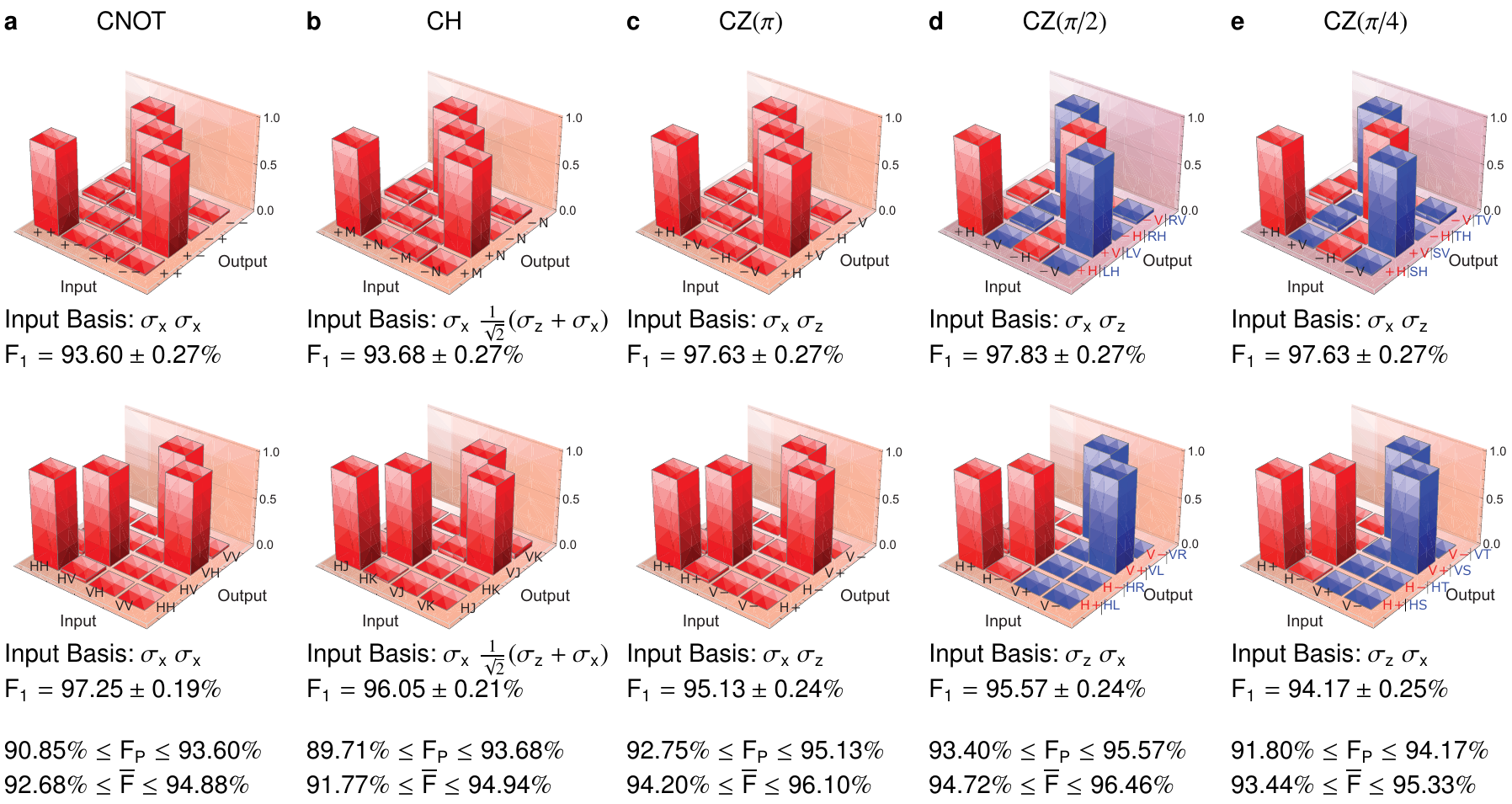}
\end{center}
\par
\vspace{-0.5cm}
\caption{\textbf{Implementing two-qubit controlled-unitary operations by harnessing entanglement in a larger Hilbert space.} (\textbf{a--e}) Experimentally measured ``truth tables" for several CU gates. 
For the inputs of the CH gate $\vert M\rangle$ and $\vert N\rangle$ ($\vert J\rangle$ and $\vert K\rangle$) are the eigenstates of $\frac{1}{\sqrt{2}}(\sigma_{z}+\sigma_{x})$
($\frac{1}{\sqrt{2}}(\sigma_{z}-\sigma_{x})$) with
$+1$ and $-1$ eigenvalues respectively: \emph{i.e}
$\vert M\rangle=\cos(\pi/8)\vert H\rangle+\sin(\pi/8)\vert V\rangle$;
$\vert N\rangle=\sin(\pi/8)\vert H\rangle-\cos(\pi/8)\vert V\rangle$;
$\vert J\rangle=\cos(5\pi/8)\vert H\rangle+\sin(5\pi/8)\vert V\rangle$;
$\vert K\rangle=\sin(5\pi/8)\vert H\rangle-\cos(5\pi/8)\vert V\rangle$.
Notice that two different output measurement bases are applied for each truth table of CPhase($\pi/2$) and CPhase($\pi/4$). The red columns correspond to red output labels while the blue columns correspond to blue output labels which include states such as $\vert HL\rangle$, $\vert HR\rangle$, $\vert HS\rangle$ and $\vert HT\rangle$ where $\vert R/L\rangle=\frac{1}{\sqrt{2}}(\vert H\rangle\pm i\vert V\rangle)$ and $\vert T/S\rangle=\frac{1}{\sqrt{2}}(\vert H\rangle\pm e^{i\pi/4}\vert V\rangle)$. Bounds on the process fidelities $F_P$ and average fidelities $\overline{F}$ are calculated from the classical fidelities shown under the truth tables. Each truth table requires 16 measurements and each measurement is taken 1 s. The count of each high column is around 2000. Note that for probabilities near zero values in the truth tables we increased the integration time from 1 s to 10 s in order to improve the statistics.}
\label{fig4}
\vspace{-0.5cm}
\end{figure*}

\vspace{5mm}

\noindent\textbf{\changedb{Experimental demonstration}} We now present experimental demonstrations of several two-qubit $CU$ gates using this general entanglement-based method (specifically corresponds to Figure~\ref{fig2}d). Figure~\ref{fig3} shows a schematic diagram of our experiment. Note the simplicity of this scheme relative to the previous demonstration of photonic $CU$ gates~\cite{la-nphys-5-134}, in particular the fact that it requires no quantum interference.  A continuous-wave laser is focused onto a BiBO crystal and thus produces photon pairs through the type-I spontaneous parametric-down conversion (SPDC) process. 
We collect two photons from four points from the SPDC cone, as
shown, resulting in a two-photon four-mode state
\cite{al-prl-102-153902} of the form
$\frac{1}{\sqrt{2}} (|H \rangle_{1r}|H \rangle_{2r}
+ |H\rangle_{1b} |H\rangle_{2b}) $
. 
By passing each mode through several waveplates, we prepare a state $\frac{1}{\sqrt{2}}\left(\vert\phi\rangle_{1r,2r}+\vert\phi\rangle_{1b,2b}\right)$, where $\vert\phi\rangle$ can be an arbitrary
two-qubit separable state. Here we designate photon 1 as the control, which is in modes $1r$ and $1b$, and photon 2 as the target, which is in modes $2r$ and $2b$. Before modes $1r$ ($2r$) and $1b$ ($2b$) are combined at a BS, we let the four modes pass through four single-qubit gates $A1$, $B1$, $A2$ and $B2$ which are constructed from waveplates or PBSs. Then, by measuring the two-photon coincidences between detectors at ports $1$ and $2$, we get the state $(A+B)\vert\phi\rangle$, where $A=A1\otimes A2$ and $B=B1\otimes B2$. As explained above , by setting $A1=\vert H\rangle\langle H\vert$, $A2=I$, $B1=\vert V\rangle\langle V\vert$ and $B2=U$, the corresponding $CU$ gate is obtained.

We constructed a series of $CU$ gates, including CNOT, C-Hadamard (CH), CPhase (CZ), CZ($\pi/2$) and CZ($\pi/4$) gate, by setting $B2=X, H, Z, Z_{\pi/2}$ and $Z_{\pi/4}$, respectively. To evaluate the performance of these gates, we adopted the method introduced in ref.~\onlinecite{ho-prl-94-160504}. For each gate, 
two truth tables are measured in complimentary bases. The bases we chose are shown in Figure 4.
The fidelity of these truth tables with the ideal ($F_1$ and $F_2$) bound the process fidelity of the gate via:
\begin{equation}
\left(F_1+F_2-1\right)\leq F_P\leq Min\left(F_1,F_2\right)
\end{equation}
We have also performed full process tomography on one of these gates (see Methods section).
From $F_P$ we can calculate the output state fidelity averaged over all input states via the average gate fidelity:
\begin{equation}
\overline{F}=\frac{dF_P+1}{d+1},
\end{equation}
where $d$ is the dimension of the gate ($d=4$ for a two-qubit gate).
The results are shown in Figure~\ref{fig4}.

\begin{figure*}[t]
\begin{center}
\includegraphics[width=\textwidth]{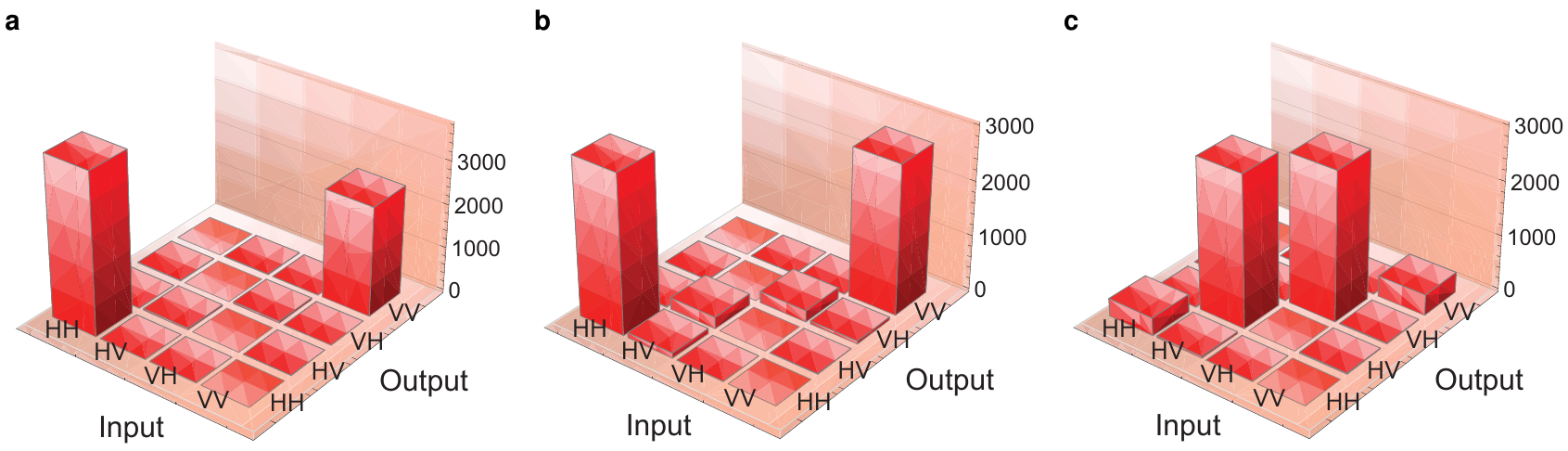}
\end{center}
\par
\vspace{-0.5cm}
\caption{\textbf{An entanglement filter and entanglement splitter. }(\textbf{a}) Logical basis truth table for an entanglement filter; the classical fidelity is $98.03\pm0.17\%$. (\textbf{b,c}) Logical truth tables for the entanglement splitter. The entanglement splitter's two outputs correspond to two complementary entanglement
 filters: one transmits $\vert HH\rangle$ and $\vert VV\rangle$ components ({b}); and the other
 transmits $\vert HV\rangle$ and $\vert VH\rangle$ components whose truth table is shown in ({c}); their classical fidelities in $H/V$ basis are $90.27\pm0.37\%$ and $87.38\pm0.42\%$ respectively.
  Here the classical fidelity is defined as the ratio of correctly transmitted photon pairs to the total number of transmitted photon pairs. The data of each column corresponds to 1 s of measurement.}
  \label{fig5}
  \vspace{-0.5cm}
 \end{figure*}

This method is not limited to realizing $CU$ gates. By changing the values of $A1$, $A2$, $B1$, and $B2$, one can implement various two-qubit quantum operations. For example, by setting $A1=A2=\vert H\rangle\langle H\vert$ and $B1=B2=\vert V\rangle\langle V\vert$, one can realize a very useful quantum gate known as entanglement filter ($EF$) \cite{ho-prl-88-147901,ok-sci-323-483}. An $EF$ is a special (non-unitary) quantum gate which filters multi-qubit states on the basis of correlations. Here, our two-photon $EF$ transmits photon pairs only if they share the same horizontal or vertical polarization, without measuring the polarization state. Compared with the previous method \cite{ho-prl-88-147901,ok-sci-323-483}, our method is simpler and more intuitive. \changedb{We implement this two-photon $EF$ using the setup shown in Figure~\ref{fig3} where $A1=A2=\vert H\rangle\langle H\vert$ and $B1=B2=\vert V\rangle\langle V\vert$ and the results are shown in Figure~\ref{fig5}a.}

Another interesting feature of our approach is that the method of realizing a quantum gate is not unique: one can choose different sets of $A$ and $B$ to get the same $A+B$. Take the two-photon $EF$ for example, we can realize it in another way by setting $A1=A2=I$ and $B1=B2=Z$. This can be verified by comparing the matrix of $I\otimes I+Z\otimes Z$ with $\vert H\rangle\langle H\vert\otimes\vert H\rangle\langle H\vert+\vert V\rangle\langle V\vert\otimes\vert V\rangle\langle V\vert$ which are equivalent up to a constant of order unity. Unlike the first way of realizing an $EF$, one does not use any projection but only unitary operators which means that in fact no components are filtered out: While the $\vert H\rangle \vert H\rangle$ and $\vert V \rangle \vert V\rangle$ components would exit at $1$ and $2$ or $1'$ and $2'$ which corresponds to realizing the quantum gate $A+B$, the $\vert H \rangle \vert V\rangle$ and $\vert V \rangle \vert H\rangle$ components would exit at $1$ and $2'$ or $1'$ and $2$ which corresponds to realizing $A-B$. As the $\vert H \rangle \vert V\rangle$ and $\vert V \rangle \vert H\rangle$ components are not filtered out in this case, we call the device an entanglement splitter ($ES$). The $ES$ operation is deterministic and the experimental data from this gate are shown in Figure~\ref{fig5}b~$\&$~c.

Interestingly, the mechanics of the $ES$ can be understood in a different way. The following equations always hold
\begin{eqnarray}
\begin{split}
Z\otimes Z(I\otimes I+Z\otimes Z)\vert\phi\rangle & =+(I\otimes I+Z\otimes Z)\vert\phi\rangle\\
Z\otimes Z(I\otimes I-Z\otimes Z)\vert\phi\rangle & =-(I\otimes I-Z\otimes Z)\vert\phi\rangle
\end{split}
 \label{S}
\end{eqnarray}
which means $(I\otimes I+Z\otimes Z)\vert\phi\rangle$ (or $(I\otimes I-Z\otimes Z)\vert\phi\rangle$) is always the eigenstate of operator $Z\otimes Z$ with eigenvalue $+1$ (or $-1$) no matter what the two-qubit state $\vert\phi\rangle$ is. Then one can deduce that $I\otimes I+Z\otimes Z$ (or $I\otimes I-Z\otimes Z$) must be a projector which projects any input state to the eigenstate of $Z\otimes Z$ with eigenvalue $+1$ (or $-1$). As our circuit realizes $I\otimes I+Z\otimes Z$ and $I\otimes I-Z\otimes Z$ simultaneously, it can be regarded as an eigenstate generator of, or eigenvalue measuring device for, operator $Z\otimes Z$. One can easily find the above reasoning would hold when replacing $Z\otimes Z$ with any two-qubit operator $W$ which fulfill $W^2=I\otimes I$. This example nicely illustrates the power of linear combination of quantum gates. Although an alternative implementation of an $EF$ has previously been reported \cite{ok-sci-323-483}, the $ES$ has not been realized; as far as we know, our method is the only solution for constructing these kinds of entangling gates.


The imperfections of our experimental results are mainly due to three effects. First, the photons generated in the SPDC source are not completely indistinguishable; second, the phase between the two spatial modes is not perfectly stabilized (to zero); third, the optical components are not perfectly set to the desired values (e.g. waveplates' angles).
For the $EF$ experiment, $A1$, $B1$, $A2$ and $B2$ are all projectors and implemented with PBS. In this case the $\vert H \rangle \vert V\rangle$ and $\vert V \rangle \vert H\rangle$ components are nearly completely filtered out and this is not affected by phase errors 
between the two spatial modes. However in the $ES$ experiment, no component is filtered out, the suppression of $\vert H \rangle \vert V\rangle$ and $\vert V \rangle \vert H\rangle$ is all based on interference, which is very sensitive to phase errors 
between the two spatial modes. The implementation of the $CU$ gates is in between of the two previous cases, where $A1$, $B1$ are projectors and $A2$, $B2$ are unitaries. This explains why the fidelity values of the $CU$ gates are higher than those of the $ES$ and lower than those of the $EF$.

\vspace{5 mm}
\noindent\textbf{\large{\changedb{Discussion}}}

\noindent
\changedb{Our} method will allow simplification of small scale linear optical circuits. 
For example, the CNOT gate (or other entangling gates) demonstrated above could be combined with a post-selected version of the same gate to perform a sequence of two entangling gates on two photonic qubits. This would require two photons rather than four --- which would normally be required for the first gate to be heralded. This type of approach is likely to be of great benefit in circuits of up to 6-10 photons where the appropriate entangled states can be generated \cite{al-prl-102-153902} and one can rely on inefficient measurement.

This new approach to realizing quantum circuits also enables a quantum state to control the implementation of a quantum gate, thereby opening up the possibility to have truly quantum inputs to quantum information processors. As shown in Fig.~\ref{fig2}a, if we set the control to be $\frac{1}{\sqrt{2}}(|0\rangle+|1\rangle)$ and let the red and blue spatial modes of the target pass through gate $O_1$ and $O_2$ instead of $I$ and $U$ respectively , where $O_1$ and $O_2$ represent two arbitrary quantum gates, the output state would be $\frac{1}{\sqrt{2}}(|0\rangle O_1|\psi\rangle+|1\rangle O_2|\psi\rangle)=\frac{1}{\sqrt{2}}(|0\rangle O_1+|1\rangle O_2)|\psi\rangle$, where $|\psi\rangle$ is the initial target state. In some sense, our circuit realizes a peculiar entangled ``state" $\frac{1}{\sqrt{2}}(|0\rangle O_1+|1\rangle O_2)$ in which a quantum bit and a quantum gate are entangled (without needing to know what the gate is). This peculiar entanglement  connecting a qubit with quantum gates may have some useful implications. An immediate application is to teleport a qubit onto a quantum gate by using this entangled ``state". In this sense, one can get a quantum gate $\alpha O_1+\beta O_2$ by teleporting a qubit $\alpha|0\rangle+\beta|1\rangle$. This is a novel way to control a quantum gate by using a quantum bit that is related to ideas of programmable quantum gate arrays \cite{ni-prl-79-321}.

In summary, we have proposed a different approach to realizing the controlled operations that are at the heart of the majority of important quantum algorithms. \changedb{With} this method, one can directly integrate an arbitrary \changedb{operation} into the circuit to build the corresponding $CU$ gate even if the unitary $U$ is unknown. This \changedb{is} in contrast to other methods that harness extra degrees of freedom \cite{ra-pra-75-022313,la-nphys-5-134} that work only for known single target qubit unitaries. Our method is not limited to $CU$ gates but can be extended to realize more general entangling gates. 
We demonstrated the power of this approach by experimentally implementing several high-fidelity two-qubit gates. In each case the implementation of the control circuit was completely independent of the choice of quantum operation. This method has the potential to change the way we implement quantum circuits for all algorithms and will find a wide range of applications across quantum information science and technology as the complexity of the quantum circuits implemented grows to include more sophisticated algorithms.

\vspace{6pt}

\vspace{5 mm}
\noindent\textbf{\large{\changedb{Methods}}}

\begin{figure*}
\begin{center}
\includegraphics[width=0.9\textwidth]{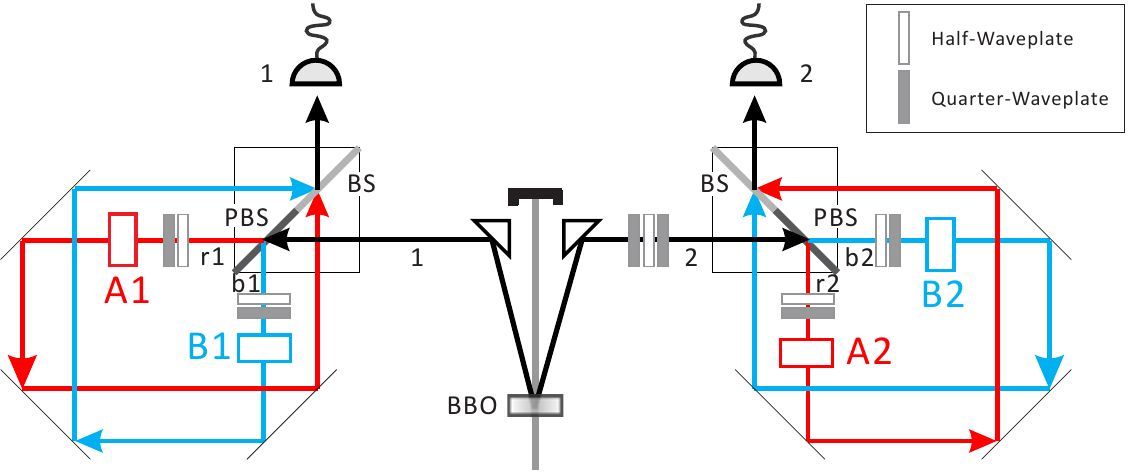}
\vspace{-0.5cm}
\end{center}
\caption{\textbf{Experimental setup in a Sagnac structure.} A \changedb{60 mW} continuous-wave (CW) laser beam with a central wavelength of \changedb{404}nm is focused \changedb{onto} a type-II BBO crystal to create entangled photon-pairs. The PBS part of the cubes convert the polarization-entangled state to spatial-entangled state. $A1$, $B1$, $A2$ and $B2$ are four single-qubit gates. By post-selecting the case where the two photons exit at ports $1$ and $2$, one would effectively realize a two-qubit quantum gate $(A1\otimes A2+B1\otimes B2)$. The displaced-Sagnac structure makes the phase between modes $r1$ and $b1$ ($r2$ and $b2$) inherently stable.
}
\label{fig6}
\vspace{-0.5cm}
\end{figure*}

\small{\noindent\textbf{\changedb{An alternative experimental demonstration}} In the experiments described in the main text, we prepared the spatial entangled state $\frac{1}{\sqrt{2}}\left(\vert\phi\rangle_{1r,2r}+\vert\phi\rangle_{1b,2b}\right)$ as the input~(see Figure 3) for implementing the various two-qubit quantum gates, where $|\phi \rangle$ is an arbitrary polarization-encoded two-qubit separable state. Here we want to point out that the phase between the two components $\vert\phi\rangle_{1r,2r}$ and $\vert\phi\rangle_{1b,2b}$ was stabilized by using monitoring and feedback method. Although the phase stabilizing approach is good enough to construct these quantum gates, for application such as phase estimation algorithm, where phase itself is the target to be measured, a setup with inherent phase stability is required.}

\small{Here we present an experimental demonstration of the same scheme by using a setup which is inherently phase stable. Instead of using type-I spontaneous parametric down-conversion (SPDC) source to get the spatial entangled photon pairs, we use type-II SPDC source to get polarization entangled photon pairs first and then convert them to the spatial entangled ones. In this way, we can build a displaced-Sagnac structure in the setup to make the phase stable. Figure~\ref{fig6} shows the schematic diagram of our experiment. We use the same continuous wave laser to pump a BBO cystal cut for type-II SPDC and get the two photon state $\frac{1}{\sqrt{2}} (|H \rangle_{1}|V \rangle_{2}+ |V\rangle_{1} |H\rangle_{2})$.}

\begin{figure*}
\vspace{0.2cm}
\begin{center}
\includegraphics[width=.84\textwidth]{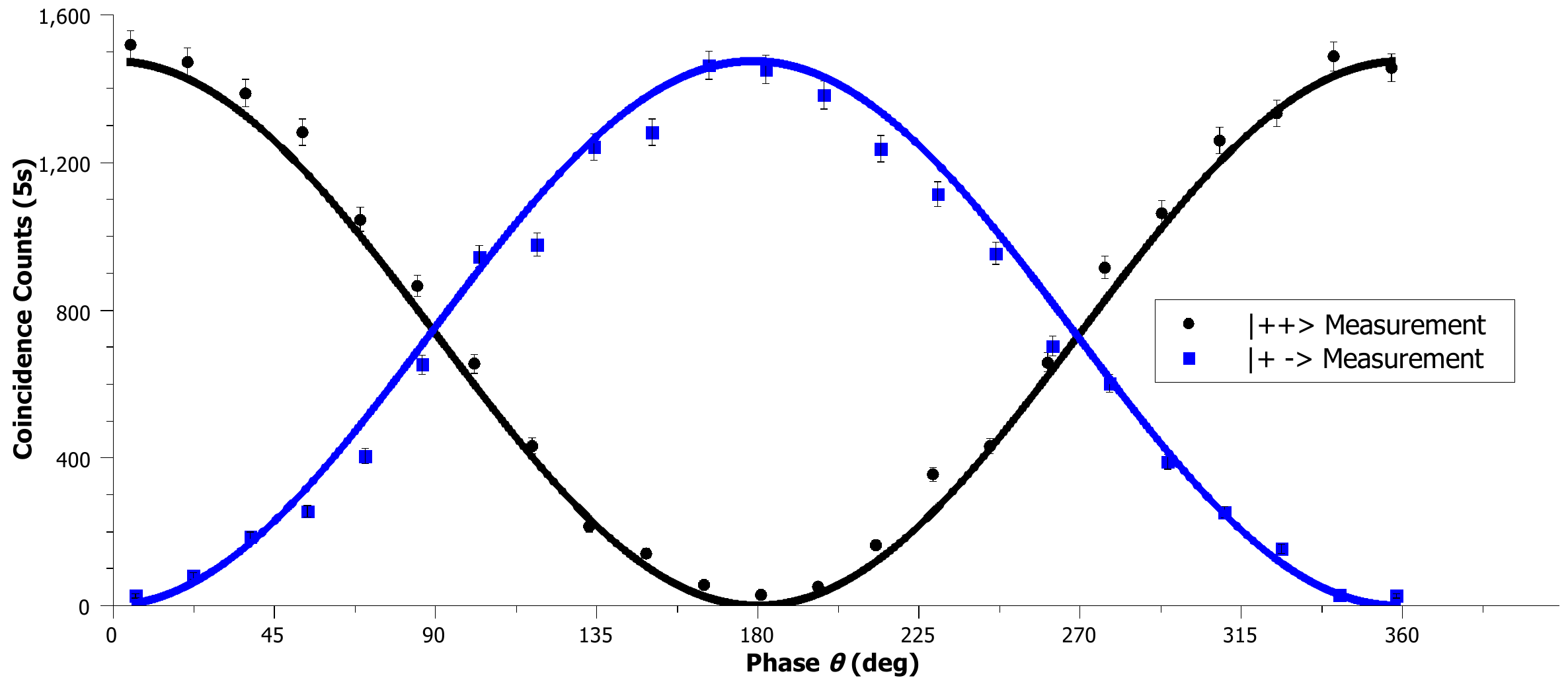}
\vspace{-0.5cm}
\end{center}
\caption{\textbf{Coincidence rates as a function of the phase between spatial modes r and b.} The two-photon output state would be $\frac{1}{\sqrt{2}} (|H \rangle_{1r}|V \rangle_{2r}+ e^{i\theta}|V\rangle_{1b} |H\rangle_{2b})$, where $\theta$ is the relative phase between spatial modes r and b. The coincidence rates of $\vert+\rangle\vert+\rangle$ (or $\vert+\rangle\vert-\rangle$) would be $C(1+\cos\theta)/2$ (or $C(1-\cos\theta)/2$) where C is the maximum count rate of $\vert+\rangle\vert+\rangle$ (or $\vert+\rangle\vert-\rangle$).
}
\label{fig7}
\vspace{-0.5cm}
\end{figure*}

\begin{figure*}
\begin{center}
\includegraphics[width=\textwidth]{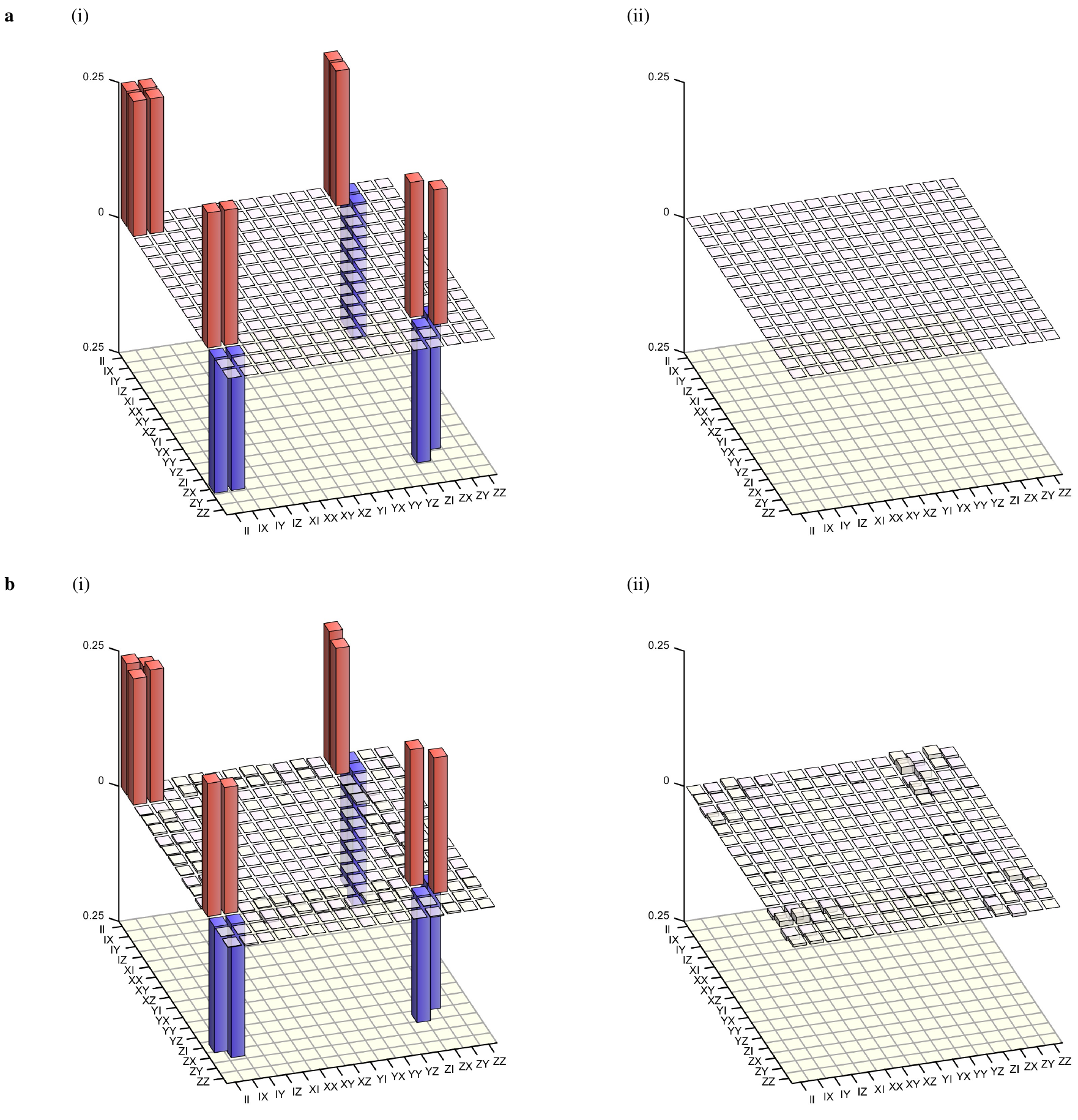}
\end{center}
\par
\vspace{-0.5cm}
\caption{\textbf{Process matrix of the CNOT gate} (\textbf{a}) Ideal process matrix. (\textbf{b}) Maximum-likelihood experimental reconstructed process matrix. (i) Real and (ii) imaginary parts are shown. We observe a high process fidelity of $96.13\pm0.17\%$ with the ideal case. The error estimate is obtained by performing many reconstructions with random noise added to the raw data in each case. Matrices are presented in the standard Pauli basis. \cite{ob-prl-93-080502}
}
\label{fig8}
\vspace{-0.5cm}
\end{figure*}

\small{By placing PBS/BS cubes (half PBS, half BS) on both arms and letting the photons pass through the PBS part of the cubes, we would get the state $\frac{1}{\sqrt{2}} (|H \rangle_{1r}|V \rangle_{2r}+ |V\rangle_{1b} |H\rangle_{2b})$. The HWP and QWP on each path further convert the state to $\frac{1}{\sqrt{2}}\left(\vert\phi\rangle_{1r,2r}+\vert\phi\rangle_{1b,2b}\right)$ which is exactly same as the spatial entangled state in the original experiments.
We let the four spatial modes $1r$, $2r$, $1b$ and $2b$ pass through four single-qubit gates $A1$, $A2$, $B1$ and $B2$ respectively and then mix the spatial modes $1r$ and $2r$ ($1b$ and $2b$) on the BS part of the cube. By postselecting the case when two photons exit at ports $1$ and $2$, the quantum operation $A+B$ is obtained. Here we set $A1=\vert H\rangle\langle H\vert$, $B1=\vert V\rangle\langle V\vert$, $A2=I$, $B2=X$ and thus realize a CNOT gate.}

\small{As the phase $\theta$ between $r$ and $b$ modes is inherently stable in our setup, we can now show the variation of  the coincidence rates with the phase $\theta$ change.  By setting $|\phi \rangle$ to be $\left\vert +\right\rangle\left\vert H\right\rangle$, the state $\frac{1}{\sqrt{2}}(|H\rangle|H\rangle+|V\rangle|V\rangle)$ would be obtained. We then place a set of waveplates (QWP, HWP and QWP) in the path of photon $2$ as shown to continuously tune the phase between the spatial modes $r$ and $b$ and measure the $\vert+\rangle\vert+\rangle$ and $\vert+\rangle\vert-\rangle$ two-photon coincidence rates at the output, where $\vert\pm\rangle=\frac{1}{\sqrt{2}}(|H\rangle\pm|V\rangle)$. As shown in Figure~\ref{fig7}, two complementary cosine curves are obtained as expected.}

\small{We fully characterize the CNOT gate through quantum process tomography (QPT) and the experimentally reconstructed process matrices are shown in Figure~\ref{fig8}. We observe a high process fidelity of  $96.13\pm0.17\%$ with the ideal case.}

\vspace{5 mm}
\noindent\textbf{\large{\changedb{Acknowledgments}}}

\small{\noindent We thank P. J. Shadbolt for writing the code for reconstructing the process matrix and H. F. Hofmann, A. Laing, J. C. F. Matthews, A. Politi for helpful discussions. This work was supported by ERC, EPSRC, PHORBITECH, NOKIA, IARPA, the Leverhulme Trust, and NSQI. J.L.O'B. acknowledges a Royal Society Wolfson Merit Award.}






\end{document}